\documentclass[manuscript]{aastex631}
\usepackage{natbib}
\newcommand\asec{\mbox{$^{\prime\prime}$}}%
\newcommand\n{\noindent}
\newcommand\mn{\medskip\noindent}
\newcommand\bn{\bigskip\noindent}

\begin{document}

\title{Imaging Spectropolarimetry -- A New Observing Mode on the Hubble Space Telescope's Advanced
Camera for Surveys}

\author[0000-0001-6145-5090]{Nimish P. Hathi}
\affiliation{Space Telescope Science Institute \\
3700 San Martin Drive, Baltimore, MD 21218, USA}

\author[0000-0003-4653-6161]{Dean C. Hines}
\affiliation{Space Telescope Science Institute \\
3700 San Martin Drive, Baltimore, MD 21218, USA}

\author{Yotam Cohen}
\affiliation{Space Telescope Science Institute \\
3700 San Martin Drive, Baltimore, MD 21218, USA}

\author[0000-0001-9440-8872]{Norman A. Grogin}
\affiliation{Space Telescope Science Institute \\
3700 San Martin Drive, Baltimore, MD 21218, USA}

\author{Marco Chiaberge}
\affiliation{Space Telescope Science Institute \\
3700 San Martin Drive, Baltimore, MD 21218, USA}



\begin{abstract}

Imaging spectropolarimetry is a new observing mode on the Advanced Camera for Surveys (ACS) aboard the Hubble Space Telescope (HST) that was commissioned in Cycle 30 and is available to HST observers starting in Cycle 31 (i.e., from 2023). It is a technique that is accessible from ground-based observatories, but the superb spatial resolution afforded by HST/ACS combined with the slitless nature of HST/ACS grism spectroscopy opens up the possibility of studying polarized extended emission in a way that is not currently possible even with Adaptive Optics facilities on the ground. This mode could help to study interesting targets including (but not limited to) QSOs, AGN and Radio Galaxies, ISM Dust Properties, Pre-Planetary Nebulae, Proto-Planetary and Debris Disks, Supernovae/Supernova Remnants, and Solar System objects. This research note presents the preliminary results from the calibration programs used to calibrate imaging spectropolarimetry on HST/ACS.

\end{abstract}



\section{Introduction} \label{sec:intro}

\n The polarization state of light observed from astrophysical sources can provide information otherwise unobtainable from intensity alone. Examples of such information include (but are not limited to): The origin of emission (direct vs. scattered); Nature of emitting or scattering particles (e.g., relativistic electrons, small dust particles); Geometry, orientation, and detailed structure of an object; Magnetic field strengths and directions; and Scattering media properties such as sizes, morphologies, and complex refractive indices. Measurement of the polarization state of the light as a function of wavelength, known as spectropolarimetry, can provide new and valuable constraints on the nature of the light source.

\mn HST/ACS offers two sets of polarizers, one optimized for near-UV/blue wavelengths (POLUV) and one optimized for visible/red wavelengths (POLV). Each set of HST/ACS polarizers comprises three polarizing filters with relative position angles of 0\degr, 60\degr, and 120\degr. 
The POLUV set can be used throughout the UV-visible region and its useful wavelength range is approximately 2000\AA\ to 7500\AA. The POLV set is optimized for the visible spectrum region and is fully effective from 4500\AA\ to about 7500\AA. The performance degrades at wavelengths longer than about 7500\AA, but useful observations might still be obtained up to approximately 8000\AA\ but with extremely careful calibration which currently is not supported by STScI. In such cases, imperfect rejection of orthogonally polarized light must be considered during data analysis (details in \citealt{ryon23,hath24}).

\mn The visible polarizers (POL0V, POL60V, and POL120V) can be paired with the G800L grism to obtain imaging spectropolarimetry with spectral resolving power (R$\sim$100 @ 8000\AA) from $\sim$5500\AA\ to 8000\AA. Even though the wavelength range of the G800L extends to 10500\AA, the polarizing efficiency degrades rapidly for wavelengths $\ge$8000\AA, rendering them ineffective at analyzing polarized light. It is worth noting that the near-UV/blue polarizers (POLUV set) can only be paired with filters on the opposing HST/ACS filter-wheel, and thus cannot be paired with the grism.

\begin{figure}
    \centering
    \includegraphics[scale=0.395]{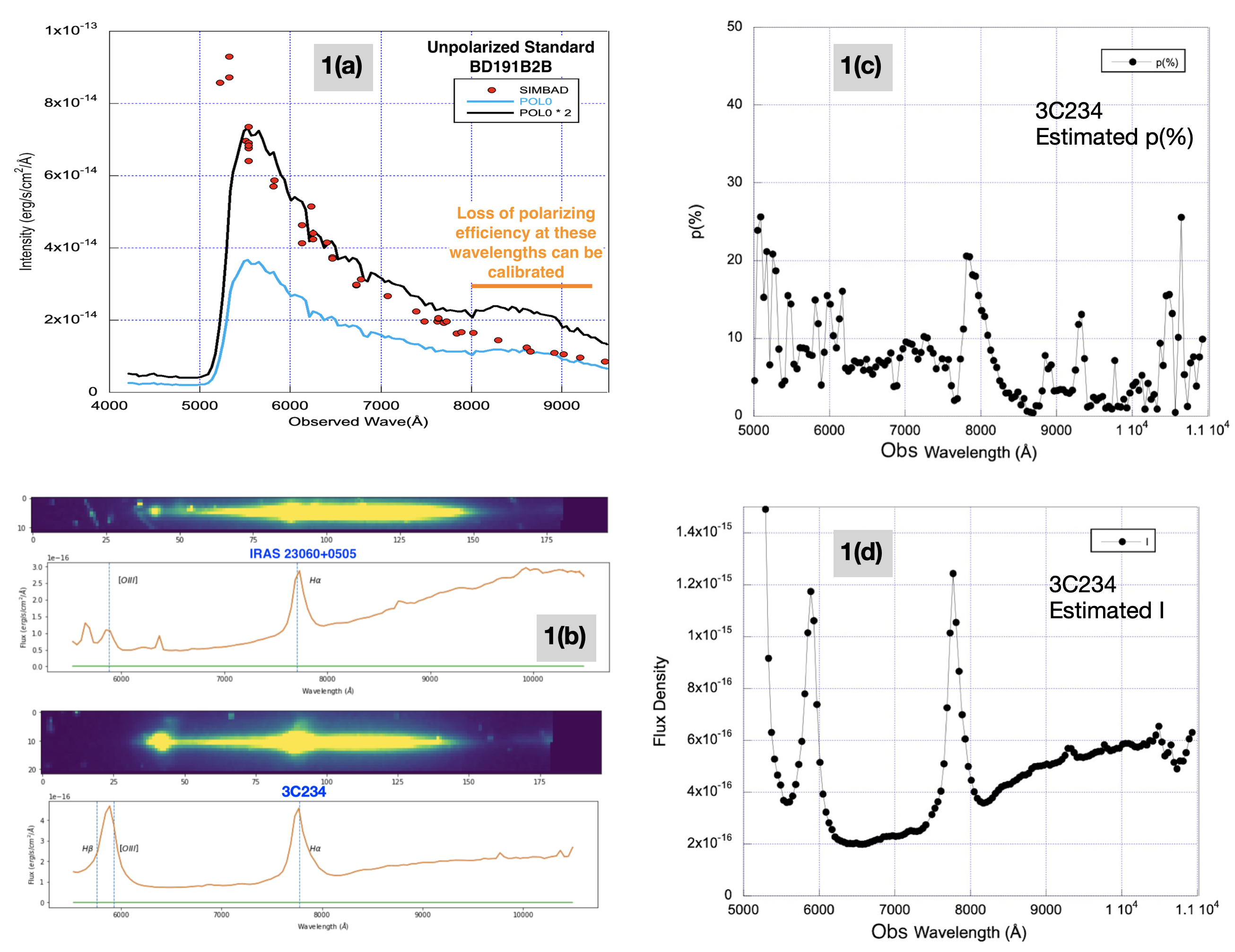}
    \caption{1(a): The spectropolarimetry data for the unpolarized standard (G191-B2B) is shown in comparison with the SIMBAD \citep{weng00} data points; 1(b): The 2D and 1D extracted spectra of the primary polarized calibration sources IRAS 23060+0505 and 3C234 galaxies; 1(c) and 1(d): A preliminary estimate of the polarization parameters -- degree of polarization, $p$ and flux density, $I$ -- for 3C234.}
    \label{fig:enter-label}
\end{figure}

\section{Calibration Programs} \label{sec:calib}

\n To calibrate the newly commissioned imaging spectropolarimetry observing mode on HST/ACS, we have collected data for 3 targets in HST Cycles 28--30 through 3 calibration programs (PID: 16474, 16869, 17257; PI Hines). One unpolarized star and two slightly different polarized calibrators were observed to investigate the bluer ($\sim$6500\AA) and redder ($>$7500\AA) sensitivity of polarizers.

\begin{itemize}
    \item BD+52-913 (G191-B2B/BD191B2B) $\longrightarrow$ unpolarized white dwarf star [Figure 1(a)]
    \begin{itemize}\addtolength{\itemsep}{-0.25\baselineskip}
        \item a null calibrator
        \item used to estimate instrumental polarization 
    \end{itemize}
    \item IRAS 23060+0505 $\longrightarrow$ polarized hyper-luminous infrared galaxy [Figure 1(b)]
    \begin{itemize}\addtolength{\itemsep}{-0.25\baselineskip}
        \item a polarized calibrator (5--20\% polarized)
        \item $z=0.1738$, H$\alpha$ at $\sim$7700\AA
        \item polarization at bluer wavelengths ($\leq$7500Å) is 5--10\% but higher (10--20\%) at redder ($>$7500\AA) wavelengths so this target is primarily used to investigate the red sensitivity
    \end{itemize}
    \item 3C234 $\longrightarrow$ radio galaxy with polarized broad emission lines [Figure 1(b)]
    \begin{itemize}\addtolength{\itemsep}{-0.25\baselineskip}
        \item a polarized calibrator (10--15\% polarized)
        \item $z=0.1848$, H$\alpha$ at $\sim$7775\AA
        \item polarization at bluer wavelengths ($\leq$7500Å) is $\ge$10\% so this target is primarily used to investigate the blue sensitivity but it also helps to double-check the red sensitivity
    \end{itemize}    
\end{itemize}

\section{Preliminary Results} \label{sec:prelim}

\n Figures 1(a), 1(c) and 1(d) show initial results/comparisons of an unpolarized star (G191-B2B/BD191B2B) and a polarized radio galaxy (3C234). Our preliminary results are consistent with the \citealt{tran95} Keck spectropolarimetry observations. Commissioning and characterization of this observing mode indicates that this new mode will be capable of measuring polarization signals with precision in percentage polarization $\sim$1--2\%, and with similar absolute accuracy. The instrumental polarization is estimated at $<$2--4\% and is calibrated out during data processing. Because this observing mode is slitless, it is most suited for point sources in uncrowded fields. However, slightly extended sources up to 2--3\asec\ have been observed as polarization calibrators during the commissioning process, and reliable results should still be obtained for such objects with intrinsic polarization $\ge$4--8\%. Note that polarization follows a Rice as opposed to a Poisson distribution, so confident polarization measurements require an absolute minimum $p/\sigma_p$ $\ge$ 4, where $p$ is the fractional polarization/degree of polarization and $\sigma_p$ is the uncertainty in the fractional polarization/degree of polarization. However, $p/\sigma_p$ $\ge$ 5 is highly recommended. If necessary, pixels may be binned to achieve this requirement, but at the expense of spatial and spectral resolution.

\mn As for all slitless grism observations, observers should take care in selecting a spacecraft orientation to eliminate (or at least minimize) contaminating spectra from nearby sources. In principle, observations taken at a different orientation from a first epoch can be used to disentangle the contamination, but polarimetry is very sensitive to flux mismatches, and such a procedure has not been evaluated for HST/ACS spectropolarimetry yet. 

\section{Ongoing and Future Work} \label{sec:ongoing}

We are currently working on an automated pipeline to reduce/analyze the HST/ACS imaging spectropolarimetry data. For each triple set of observations (G800L+POL0V, G800L+POL60V, G800L+POL120V), the future pipeline will perform the following steps. 

\begin{itemize}
    \item Apply distortion corrections — done with AstroDrizzle\footnote{\href{https://drizzlepac.readthedocs.io/en/deployment/astrodrizzle.html}{https://drizzlepac.readthedocs.io/en/deployment/astrodrizzle.html}} with EXP weighting. The resulting images must be in units of DN/s.
    \item Align images – this alignment/registration needs to be more accurate than just using the WCS. A zeroth-order or a star in the field can be used.
    \item Form the Stokes parameters, $Q$ \& $U$, and also the fractional polarization ($p$) and position angle on the sky/the polarization angle ($\theta$) using the prescription shown in Section 5.3 of the HST/ACS Data Handbook \citep{hath24}, but without applying the transmission correction term
    \item Derive and apply the polarization efficiency correction
    \item Extract $Q$, $U$, $p$, and $\theta$ spectra as needed using the standard HST/ACS grism extraction software, currently, HSTaXe \citep{sose23}
\end{itemize}

\n It is crucial for proper polarimetry reduction that these steps be performed in 2D space until the very end when a 1D spectrum can be extracted (if that is the end goal).

\n In some circumstances, observers may wish to combine data from multiple visits. This may happen when observers use a different roll angle on the telescope to help separate overlapped grism spectra. In these cases, the co-additions must be performed in $Q$ \& $U$ space, not in $p$ and $\theta$ space. The co-addition in $p$-space will yield erroneous results as $p$ follows the Rice distribution, and not a Poisson distribution. Details about the pipeline and rotational translation of Stokes parameters will be published in an upcoming Instrument Science Report (ISR) as well as in the updated version of the HST/ACS Data Handbook \citep{hath24}.

\mn Observers should examine the 2D images in $Q$ \& $U$ (or in $p$ \& $\theta$) to look for stars in the field. Most stars in the galaxy have low or zero polarizations. Therefore, the field stars should all be unpolarized and have zero $Q$, $U$, and $p$. Polarization signals in field stars may indicate calibration problems. Contact the HST Help Desk (\href{https://stsci.service-now.com/hst}{https://stsci.service-now.com/hst}) for guidance unless you are aware that the field stars are polarized.

\mn The HST/ACS Instrument Team emphasizes that this is a newly commissioned mode, and is still being fully characterized and calibrated. Observers using this mode should contact the HST Help Desk (\href{https://stsci.service-now.com/hst}{https://stsci.service-now.com/hst}) to initiate a dialog with the team, for details and updates refer to the latest version of the HST/ACS Instrument Handbook \citep{ryon23} and HST/ACS Data Handbook \citep{hath24}.

\bn DATA DOI: Some/all of the data presented in this article were obtained from the Mikulski Archive for Space Telescopes (MAST) at the Space Telescope Science Institute. The specific observations analyzed can be accessed via \dataset[DOI: 10.17909/khze-hp39]{https://doi.org/10.17909/khze-hp39}. 





\end{document}